# Scanning-probe quantum sensing of microwave and static magnetic field response of an on-chip superconducting resonator

Senlei Li[1], Jingcheng Zhou[1], Zelong Xiong[1], Hanyi Lu[2,1], and Hailong Wang[1*]

[1]School of Physics, Georgia Institute of Technology, Atlanta, Georgia 30332, USA
[2]Department of Physics, University of California, San Diego, La Jolla, California 92093, USA

[*]Corresponding author: hwang3021@gatech.edu

**Abstract**: Superconducting resonators are finding increasing applications in designing advanced quantum circuits for ongoing sensing, metrology, and computing technological revolution. A detailed knowledge of microscopic electromagnetic properties of superconducting resonators is directly relevant for their further improvements on circuitry design and device performance. Here, we introduce scanning-probe quantum microscopy to report nanoscale sensing of microwave and static magnetic field environment of an on-chip niobium (Nb) superconducting resonator. Taking advantage of Rabi oscillation measurements, we show that microwave magnetic fields generated by the superconducting resonator mode can be utilized to achieve coherent control of a quantum spin sensor. We further visualize static electromagnetic field response of the Nb resonator, showing magnetic field-induced formation, evolution, and depinning of superconducting vortices. Our results provide insights into future design, testing and evaluation of solid-state superconducting resonators, highlighting the potential of quantum sensors as a local probe to investigate electromagnetic properties of superconducting quantum circuits.

Superconducting devices built on near-zero electrical resistance, quantum interference effects and unconventional electromagnetic properties serve as a leading contender for advancing the rapidly growing quantum technological revolution.[1,2] To date, they have been playing an active role in developing cutting-edge quantum sensing, quantum metrology, quantum computing and quantum simulation technologies that outperform the conventional counterparts.[1,3–6] Superconducting resonators are naturally relevant in this context thanks to their direct applications in quantum information science (QIS) such as coherent control of superconducting qubits,[7] dispersive qubit readout,[8] quantum bus transmission and others.[9,10] From a classic electromagnetic perspective, a superconducting resonator could be viewed as an *LC* circuit with a characteristic resonant frequency, low-energy dissipation, nonlinear response and kinetic inductance effect, which finds a range of applications in designing sensitive photon detectors, radiation amplifiers, filters and other advanced microwave circuits.[11–15]

In the current state of the art, the performance of a superconducting resonator is typically evaluated by microwave measurements and theoretical simulations, unable to reveal detailed information on the local electromagnetic behaviors in a real material/device environment. This certainly hinders future improvements in quality factor, electromagnetic tunability and on-chip integration with other solid-state quantum device units to realize innovative QIS circuits. In this letter, we report scanning-probe quantum imaging of nanoscale microwave and static magnetic field environment of an on-chip superconductor niobium (Nb) resonator. We show that coherent magnetic fields generated by the superconducting resonator mode can excite Rabi oscillations of a single nitrogen-vacancy (NV) center, where the driving efficiency reaches the maximum when the NV spin energy matches the resonator frequency. By performing scanning NV Rabi measurements, we have imaged spatially varying microwave field distribution in the Nb resonator. We further visualize static electromagnetic field response of the resonator, revealing microscopic formation, evolution and magnetic field-induced depinning of superconducting vortices in the Nb microwave center strip.

We first discuss the device structure and our quantum sensing platform. Figure 1(a) shows an optical microscopy image of an on-chip Nb superconducting circuit. It consists of a coplanar microwave feedline capacitively coupled with a quarter-wave resonator lithographically patterned on a sapphire ($Al_2O_3$) substrate. The width of the patterned coplanar center strip is ~20 μm and the spacing gap between the central and ground conductors is ~8 μm. The length of the microwave center strip of the resonator is 12 mm with one end open-circuited and the other end shorted, giving a theoretically calculated resonator frequency of the fundamental mode $f_{\mathrm{R}} = c/4l\sqrt{\epsilon_{\mathrm{eff}}} \approx 2.7$ GHz, where $c$ is the speed of light, $l$ is the length of the center strip, and $\epsilon_{\mathrm{eff}}$ represents the average dielectric constant of vacuum and $Al_2O_3$ substrate.[16,17] Figure 1(b) shows the microwave transmission coefficient $S_{21}$ of the patterned Nb resonator measured as a function of input microwave frequency $f$ at 5 K. The resonator frequency $f_{\mathrm{R}}$ is experimentally characterized to be ~2.715 GHz in absence of external magnetic field application, in agreement with our theoretical estimation. The quality factor $Q$ is obtained to be 2,800 by fitting the resonator mode spectrum.

We utilize scanning-probe NV microscopy to perform quantum sensing of microscopic electromagnetic response of the on-chip resonator as illustrated in Fig. 1(c). A single NV center is contained in a nanopillar of a diamond cantilever that is glued to a quartz tuning fork for force-feedback atomic force microscopy (AFM) operations.[18–28] The ultimate spatial resolution of our scanning-probe microscopy is determined by the vertical NV-to-sample distance, which is tunable from 100 nm to 500 nm in the current study. Quantum sensing measurements take advantage of an isolated $S = 1$ NV electron spin to detect dynamic and static magnetic fields.[18,20] Our discussion starts with NV sensing of microwave magnetic fields generated from the on-chip Nb resonator. When a microwave magnetic field perpendicular to the NV spin direction is applied at the electron spin resonance (ESR) frequency, an NV spin will periodically oscillate between two different states in the rotation frame, which is typically referred to as Rabi oscillations[31] as illustrated in Fig. 1(d). The amplitude of the applied microwave magnetic field transverse to the NV axis can be deduced from the Rabi oscillation frequency $f_{\mathrm{Rabi}}$.[31,32] Here, we implement NV Rabi measurements to detect coherent magnetic fields generated by the center strip of the Nb resonator. Our quantum sensing studies focus on the device area close to the shorted end of the Nb center strip, where magnitude of the microwave magnetic field is expected to reach a maximum for the convenience of driving NV Rabi dynamics.

The top panel of Fig. 1(e) shows the pulsed optical and microwave sequence for optically detected NV Rabi oscillation measurements. We first apply a 1.5-μs-long green laser pulse to initialize the NV spin to the $m_s = 0$ state. Then, a microwave pulse at the NV ESR frequency is applied in the feedline to drive NV spin transitions followed by a second green laser pulse to measure spin-dependent NV photoluminescence (PL) and to re-initialize the NV spin. The microwave pulse is applied ~500 ns after turning off the initialization green laser pulse to minimize laser-induced Joule heating effect. Time duration of the microwave pulses varies from zero to a few hundred nanoseconds to detect delay time dependent variations of NV PL. It is instructive to note that the $m_s = \pm 1$ excited NV spin states exhibit a reduced PL intensity, since they are more likely to be trapped by a non-radiative pathway through an intersystem crossing and back to the $m_s = 0$ ground state.[31,33] The bottom panel of Fig. 1(e) presents three representative NV Rabi spectra measured at 6 K. When the lower transition of NV ESR frequency $f_-$ (tunable by the external magnetic field strength) is off from the resonator frequency $f_{\mathrm{R}}$ by 61 MHz, the NV PL spectrum slowly oscillates at a Rabi frequency $f_{\mathrm{Rabi}}$ of 2.6 MHz. When $f_- - f_{\mathrm{R}}$ decreases to 26 MHz, we observe a clearly accelerated oscillation behavior with an enhanced $f_{\mathrm{Rabi}}$ of 10.1 MHz. $f_{\mathrm{Rabi}}$ further increases to a peak value of ~20.1 MHz when the NV spin energy matches frequency of the Nb resonator mode. The increase of $f_{\mathrm{Rabi}}$ when $f_-$ approaches $f_{\mathrm{R}}$ results from an enhanced NV Rabi driving efficiency by microwave magnetic field $B_{\mathrm{R}}$ (transverse to the NV spin axis) generated by the superconducting resonator mode. Figure 1(f) plots $f_{\mathrm{Rabi}}$ measured as a function of $f_-$. It is evident that coherent NV spin control becomes the most efficient at the experimental condition of $f_- = f_{\mathrm{R}}$. Considering our measurement geometry, $B_{\mathrm{R}}$ is estimated to be ~10 G parallel with the surface plane of the Nb center strip when $f_{\mathrm{R}} = f_-$ with an input microwave power of 10 dBm at 6 K.

Next, we perform scanning-probe NV measurements to spatially resolve microwave field distribution of the on-chip superconducting resonator. Figure 2(a) shows AFM imaging of a local ground-signal-ground (GSG) device area of the patterned Nb coplanar waveguide. Figure 2(b) plots a microwave field distribution map obtained from NV Rabi measurements at 2 K over the surveyed sample region. Note that the NV spin axis is 54° away from the out-of-plane direction (*z*-axis in the local coordinate frame), and its in-plane projection is along the Nb microwave center strip (*x*-axis in the local coordinate frame). Figure 2(c) shows three characteristic Rabi spectra recorded when the NV center is scanned to lateral positions: on top of the Nb center strip, close to the edge of the center strip, and within the gap between the signal and ground lines. The exact NV positions for individual Rabi measurements are highlighted by three color points shown in Fig. 2(b). It is evident that edges of the Nb center strip produce a significantly larger microwave magnetic field. Based on the microwave magnetic field, we perform HFSS electromagnetic simulations to reconstruct the microwave current distribution in the resonator as presented in Fig.

2(d) (see Supplemental Material section 1 for details). The directions of longitudinal (relative to the center strip) microwave current $J_x$ flowing in the signal and ground lines are opposite with each other. This can be qualitatively understood by conservation of electrical charge currents in superconducting circuit as the ground line essentially serves as a return path for currents flowing in the center strip.

After demonstrating NV Rabi sensing of the microwave magnetic field environment, we now utilize scanning-probe quantum microscopy to investigate spatially dependent static electromagnetic field response, specifically nanoscale formation and evolution of superconducting vortices, of the on-chip Nb resonator. In an intuitive physical picture, superconducting vortices can be viewed as individual magnetic flux quanta generated from circulating supercurrents around a conducting core as illustrated in Fig. 3(a). Vortices in a superconductor can be locally nucleated by defects, thermal heating, electric current flow, external magnetic fields and other kinds of external stimuli.[20,21,34–36] We first show magnetic field-cooling-induced formation of superconducting vortices in the on-chip Nb resonator. In these experiments, the resonator device is cooled from room temperature in presence of an out-of-plane external magnetic field $B_z$ (cooling field), and $B_z$ is kept in the follow-up scanning NV measurements. Static NV magnetometry takes advantage of the linear Zeeman effect of the $S = 1$ electron spin in optically detected magnetic resonance (ODMR) measurements, from which magnitude of a local magnetic field at an NV site can be deduced from split NV spin energy levels.[20,21] Figure 3(b) presents two representative ESR spectra when the NV center is scanned to positions right above and away from a superconducting vortex formed in the Nb center strip. The measurement temperature is 2 K (below the superconducting transition temperature of Nb) and $B_z$ is 2 G during the ODMR measurements. It is noticed that the on-vortex case shows a larger ESR splitting due to magnetic flux penetrating through the non-superconducting vortex. The reduced splitting of NV ESR energy in the off-vortex case is attributed to the Meissner screening effect of superconducting Nb, which partially shields the effective magnetic field experienced by the NV center.[20]

When scanning the NV sensor over the Nb center strip, we can visualize magnetic field-cooling-induced formation of superconducting vortices at the nanoscale. Figures 3(c)-3(e) present a series of magnetic field ($B_s$) map of a surveyed device area at 2 K under different cooling field. $B_z$ is initially set at 2.7 G, then decreases and changes the sign to −1.3 G, and −6.3 G. Note that the background of magnetic fields emanating from superconducting sample areas has been subtracted to highlight $B_s$ from vortices. For $B_z$ = 2.7 G, circularly shaped vortices visually stand out by exhibiting pronounced local penetrating magnetic fields [Fig. 3(c)]. When the sign of $B_z$ reverses, polarity of the magnetic field $B_s$ from vortices also changes accordingly [Figs. 3(d)-3(e)]. It is worth mentioning that the superconducting vortex density increases with the magnitude of $B_z$ while the lateral dimensions of vortices decrease. The magnetic flux generated by individual vortices is estimated to be 22.2 G μm$^2$ (see Supplemental Material section 2 for details), in qualitative agreement with the theoretical prediction[20] of one flux quantum $h/2e$. By performing autocorrelation operation on magnetic field maps, the corresponding periodic superconducting vortex patterns are presented in Figs. 3(f)-3(h). One can see that a higher magnetic cooling field results in spatially more compact superconducting vortex lattice with a reduced inter-vortex spacing as expected.

Microscopic formation of superconducting vortices can modify the local static magnetic field environment of a resonator device. Last, we investigate magnetic field-driven depinning effect of superconducting vortices in the on-chip Nb resonator. Here, we combine the ODMR and scanning-probe NV imaging measurements to evaluate variations of local static magnetic field emanating from the Nb device. Figures 4(a)-4(b) present two ODMR maps measured on an NV center positioned right above the Nb center strip at 6 K and 2 K. An external magnetic field $B_{\mathrm{ext}}$ is applied along the NV spin direction during the ODMR measurements. We observe reduced NV fluorescence when the input microwave drive frequency $f$ matches the NV ESR frequencies $f_{\pm}$. Here, $f_{+}$ and $f_{-}$ denote the ESR frequencies corresponding to NV spin transitions between the $m_s = 0$ and $m_s = \pm 1$ states, respectively. Figure 4(c) summarizes the local magnetic field $B_{\mathrm{NV}}$ measured from split NV energy as a function of the external magnetic field $B_{\mathrm{ext}}$ at 6 K and 2 K. $B_{\mathrm{NV}}$ is obtained to be $\frac{f_{+}-f_{-}}{2\gamma}$, where $\gamma$ is the gyromagnetic ratio. For $T$ = 6 K, two curved $f_{+}$ and $f_{-}$ lines emerge in the NV ESR map [Fig. 4(a)]. Due to the Meissner screening effect, $B_{\mathrm{NV}}$ experienced by the NV center is smaller than $B_{\mathrm{ext}}$ in the low-field regime as shown in Fig. 4(c), consistent with the ESR results [Fig. 3(b)]. As the increase of $B_{\mathrm{ext}}$, superconducting vortices start to nucleate in the Nb center strip, serving as a local pathway to converge penetrating magnetic fields, resulting in the field sensed by the NV center larger than $B_{\mathrm{ext}}$. The smoothly curved field-dependent variations of NV ESR frequencies $f_{\pm}$ are driven by continuous nucleation and depinning of superconducting vortices in Nb at 6 K.

When temperature decreases to 2 K, the pinning force becomes much stronger due to reduced thermal fluctuations, and individual vortices cannot continuously move as the external magnetic field increases. Vortices can only be unpinned at certain threshold magnetic fields, resulting in discrete "jumps" of measured NV ESR lines as shown in Fig. 4(b). Interestingly, we further observe certain magnetic hysteresis behavior of the NV ESR transitions during the forward and backward magnetic field sweeping processes (see Supplemental Material section 3 for details). To image the field-driven depinning of superconducting vortices at 2 K, Figs. 4(d)-4(f) present scanning NV imaging of local magnetic field maps at different values of $B_{\mathrm{ext}}$ corresponding to points "A", "B" and "C" marked in Fig. 4(c). One can see that the NV sensor (denoted by the black circle) is away from superconducting vortices in the initial state (point "A") as shown in Fig. 4(d). As $B_{\mathrm{ext}}$ increases, more superconducting vortices are formed and move towards the NV center, inducing a step jump of the local magnetic field $B_{\mathrm{NV}}$ detected at point "B" [Fig. 4(e)]. At point "C", the vortex density has dramatically increased, and superconducting vortices are basically distributed everywhere in the surveyed device region [Fig. 4(f)]. As $B_{\mathrm{ext}}$ further increases, superconducting vortices continue clustering in the Nb center strip, and the $B_{\mathrm{NV}}$ measured eventually approaches $B_{\mathrm{ext}}$.

In summary, we have demonstrated quantum sensing of nanoscale microwave and static electromagnetic properties of an on-chip superconducting resonator. We show that local dynamic magnetic fields generated by the Nb resonator mode can coherently control an NV center with a driving efficiency reaching a maximum when the resonator frequency matches NV ESR conditions. Spatially resolved static magnetic field response of the resonator is imaged by a scanning NV sensor, presenting microscopic formation and evolution of superconducting vortices at the nanoscale. We further evaluate magnetic field-induced depinning effect of superconducting vortices, showing its modification of local static field environment of the Nb resonator. Considering the dissipating nature, vortices formed in superconducting circuits could potentially introduce malignant perturbation to induce quantum decoherence of proximal qubit operations.[37,38] Our work provides a detailed knowledge on the microscopic electromagnetic profile of superconducting resonators, bringing insights into circuitry design to improve their performance on qubit readout and entanglement operations in integrated quantum circuits.[39] The cutting-edge quantum sensing approaches reported also pave the way for future design, testing and evaluation of local energy dissipation, functionalities and electromagnetic tunability of a broad range of on-chip superconducting circuits.

**Acknowledgements:** The authors are grateful to Zhaorong Gu and Mengqi Huang for help with device preparation and simulations. We acknowledge the support from the U.S. National Science Foundation under award No. ECCS-2525800.

**Author contributions:** S. L. performed quantum sensing measurements and analyzed the data. J. Z., Z. X., and H. L. contributed to the device fabrication and characterization. H. W. supervised the work.

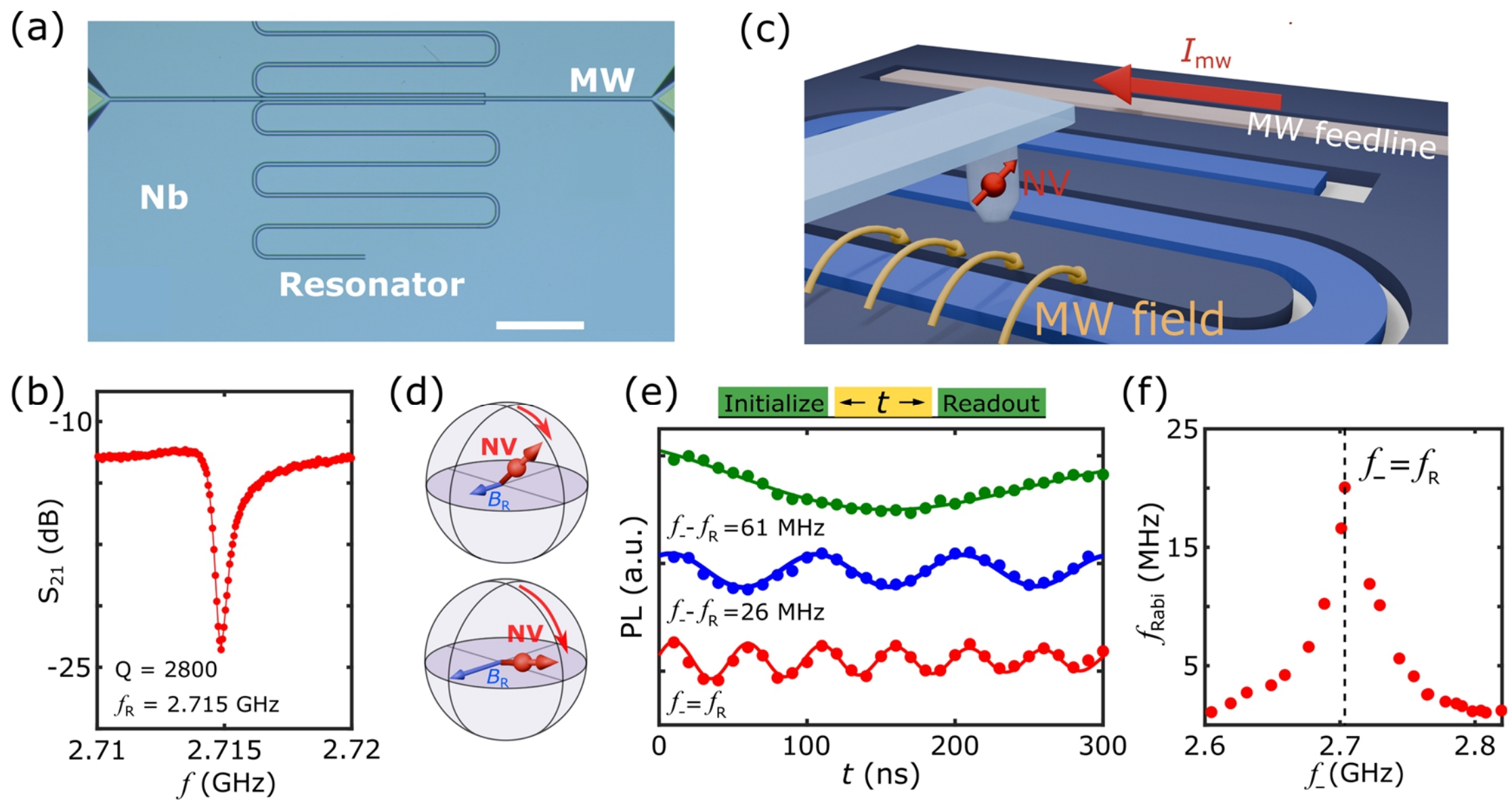

**Figure 1. NV Rabi sensing of a superconducting resonator mode.** (a) Optical image of a patterned on-chip Nb resonator. Scale bar is 500 μm. (b) Microwave transmission spectrum of a superconducting resonator mode measured at 5 K with zero external magnetic field. (c) Schematic of local NV sensing of microwave (MW) magnetic field emanating from the center strip of an on-chip Nb resonator. Microwave (MW) current $I_{\mathrm{mw}}$ flows in the feedline to drive the resonance. (d) Schematic of NV Rabi oscillations on the Bloch sphere. A larger transverse (relative to the NV spin axis) microwave magnetic field drives a faster NV spin rotation. (e) Top panel: pulsed microwave and optical sequences for NV Rabi oscillation measurements. Bottom panel: NV Rabi spectra recorded at $f_- - f_{\mathrm{R}}$ of 61 MHz, 26 MHz and 0 MHz. (f) NV Rabi frequency measured as a function of the ESR frequency $f_-$. The dashed lines highlight the condition where $f_-$ matches $f_{\mathrm{R}}$. The NV center is positioned right above the Nb center strip for Rabi oscillation measurements presented in Figs. 1(e)-1(f). The NV-to-sample distance is fixed at ~500 nm and measured temperature is 6 K for presented NV Rabi results.

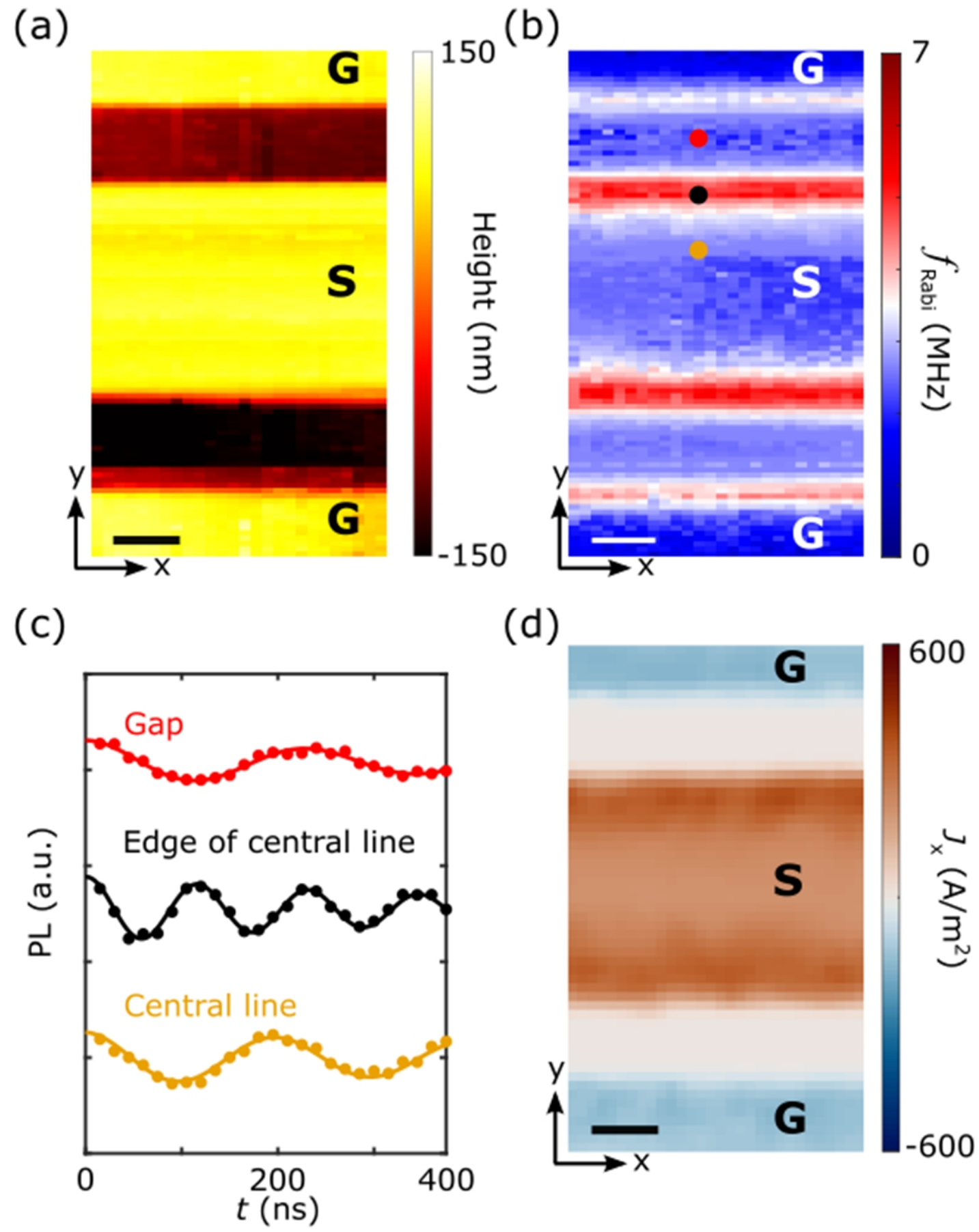

**Figure 2. Imaging microwave field distribution of an on-chip superconducting resonator.** (a) AFM imaging of a local ground-signal-ground (GSG) device area of the Nb coplanar waveguide resonator. (b) Scanning NV imaging of spatially dependent microwave magnetic field environment of the surveyed device area. The vertical NV-to-sample distance is ~500 nm, NV ESR frequency is 2.875 GHz, input microwave power is 10 dBm, and measurement temperature is 2 K. (c) Three representative NV Rabi spectra recorded when the NV center is scanned to different lateral positions highlighted by red, black, and yellow points shown in Fig. 2(b). (d) Simulated microwave current $J_x$ distribution in the Nb resonator. Scale bar is 6 μm.

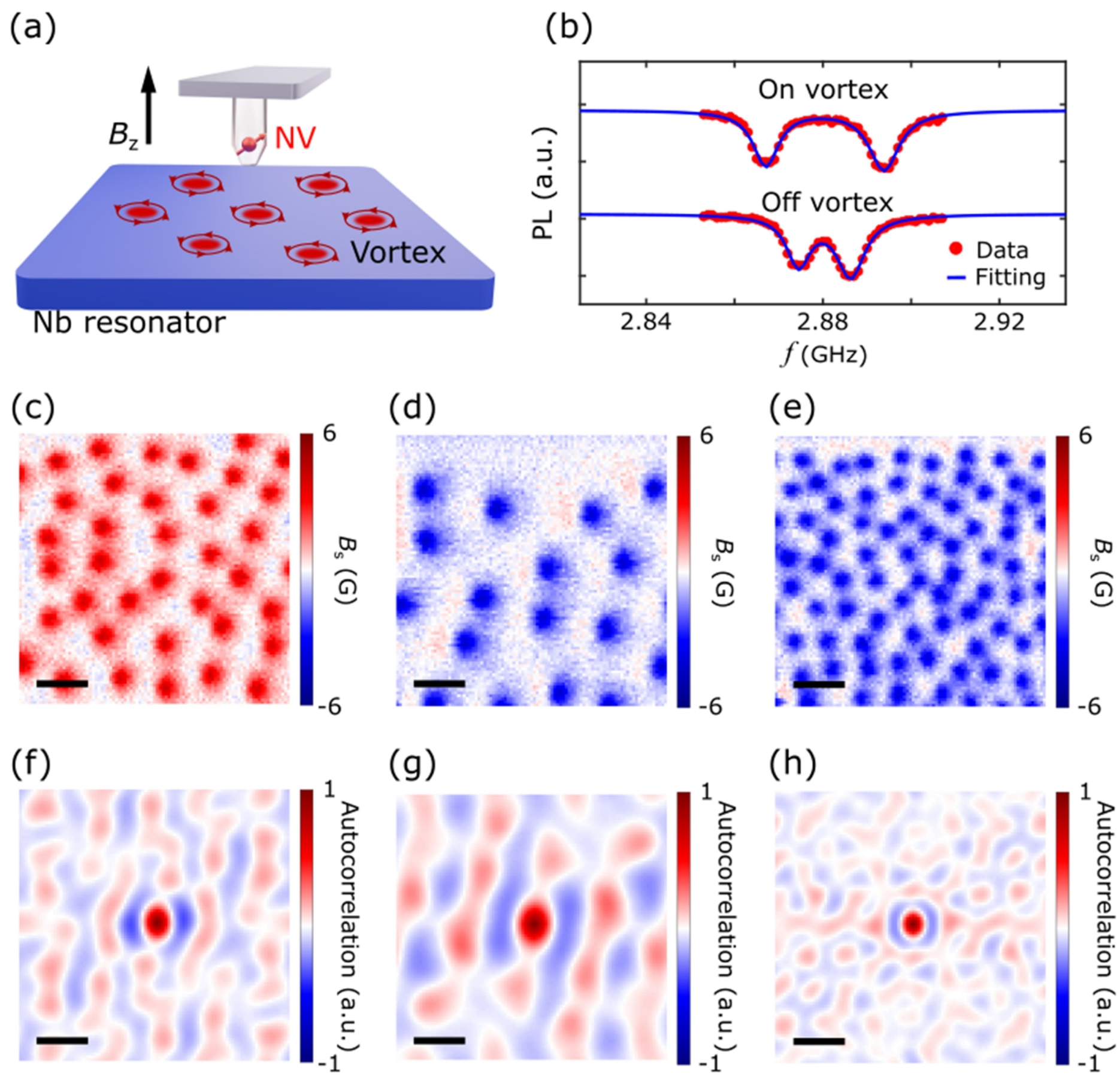

**Figure 3. Visualizing superconducting vortices formed in an on-chip Nb resonator.** (a) Schematic of imaging superconducting vortices using a scanning NV spin sensor. (b) ODMR spectra recorded when an NV center is positioned right above and away from a superconducting vortex. The NV-to-sample distance is ~100 nm for presented ODMR measurements. Double Lorentzian function is used to fit positions of NV ESR frequencies. (c)-(e) Static magnetic field ($B_s$) maps of vortices formed in the Nb resonator with a perpendicular magnetic cooling field $B_z$ of 2.7 G (c), −1.3 G (d), and −6.3 G (e). (f)-(h) Normalized autocorrelation (AC) maps of the corresponding magnetic field patterns shown Figs. 3(c)-3(e). Scale bar is 2 μm and the measurement temperature is 2 K.

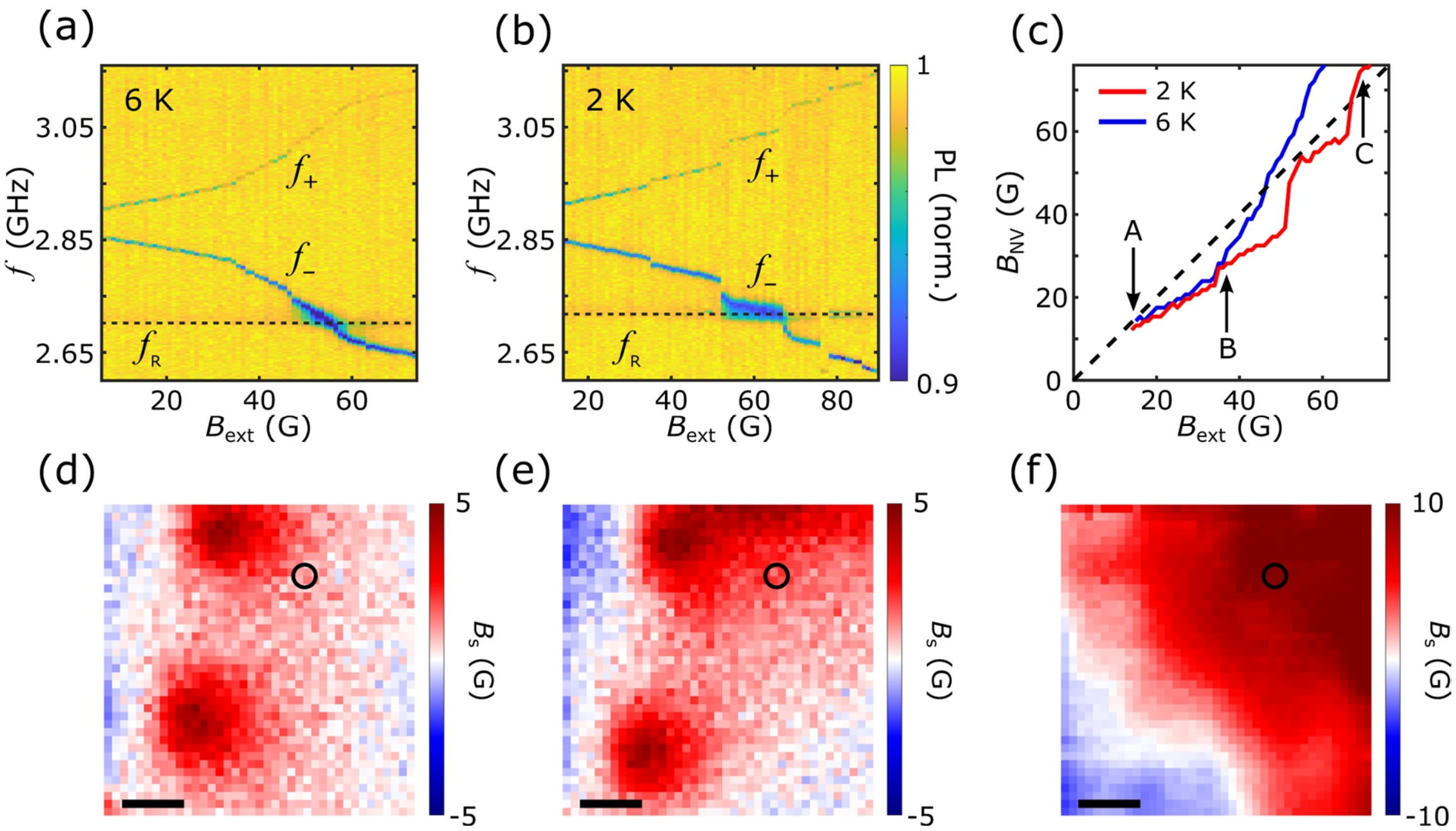

**Figure 4. Magnetic field-induced depinning effect of superconducting vortices.** (a)-(b) NV ODMR maps showing normalized PL intensity as a function of external magnetic field $B_{ext}$ applied along the NV spin direction and microwave frequency $f$ at 6 K (a) and 2 K (b), respectively. Dashed lines indicate the resonator frequency $f_R \sim 2.7$ GHz. (c) Local magnetic field $B_{NV}$ measured by the NV center as a function of $B_{ext}$ at 6 K and 2 K. Black dashed lines serve as a visual guideline of $B_{NV} = B_{ext}$. (d)-(f) Scanning-probe NV imaging of magnetic field emanating from a surveyed local device area when $B_{ext}$ = 14 G (d), 35 G (e), and 70 G (f) at 2 K, which correspond to the external magnetic field conditions of points "A", "B", and "C" marked in Fig. 4(c). The black circle denotes the lateral position of the NV sensor for ODMR measurements presented in Fig. 4(b). Scale bar is 1 μm.